\documentclass[aps,prl,twocolumn,showpacs,superscriptaddress,groupedaddress]{revtex4}
\usepackage{enumerate,url,mathrsfs}
\usepackage{latexsym}
\usepackage{amsthm}
\usepackage{amsfonts}
\usepackage{amssymb}
\usepackage{times,txfonts,yfonts}
\usepackage{hyperref}
\usepackage{color}
\usepackage{graphics}
\usepackage[titletoc,title]{appendix}
\usepackage{lipsum}
\usepackage{epsfig}
\DeclareGraphicsExtensions{.eps,.ps,.pdf}

\begin{document}

\title{Time dilation in relativistic quantum decay laws of moving unstable particles}

\author{Filippo Giraldi}
\email{giraldi@ukzn.ac.za}
\affiliation{ School of Chemistry and Physics, University of KwaZulu-Natal
\\and National Institute for Theoretical Physics (NITheP)\\
Westville Campus, Durban 4001, South Africa}

\begin{abstract}
The relativistic quantum decay laws of moving unstable particles 
are analyzed for a general class of mass distribution densities which behave as power laws near the (non-vanishing) lower bound $\mu_0$ of the mass spectrum. The survival probability $\mathcal{P}_p(t)$, the instantaneous mass $M_p(t)$ and the instantaneous decay rate $\Gamma_p(t)$ of the moving unstable particle are evaluated over short and long times for an arbitrary value $p$ of the (constant) linear momentum. The ultrarelativistic and non-relativistic limits are studied. Over long times, the survival probability $\mathcal{P}_p(t)$ is approximately related to the survival probability at rest $\mathcal{P}_0(t)$ by a scaling law. 
The scaling law can be interpreted as the effect of the relativistic time dilation if the asymptotic value $M_p\left(\infty\right)$ of the instantaneous mass is considered as the effective mass of the unstable particle over long times. The effective mass has magnitude $\mu_0$ at rest and moves with linear momentum $p$ or, equivalently, with constant velocity 
$1\Big/\sqrt{1+\mu_0^2\big/p^2}$. The instantaneous decay rate $\Gamma_p(t)$ is approximately independent of the linear momentum $p$, over long times, and, consequently, is approximately invariant by changing reference frame.
\end{abstract}

\maketitle

\section{Introduction}\label{1}
The description of the decay laws of unstable particles via quantum theory has been a central topic of research for decades 
\cite{Khalfin1997,FondaGirardiRiminiRPP1978}. Many unstable particles which are generated in astrophysical phenomena or high-energy accelerator experiments, are moving in the laboratory frame of the observer at relativistic or ultra-relativistic velocity. For this reason, plenty of studies have been devoted to formulate the decay laws in terms of relativistic quantum theory. See \cite{BakamjianPR1961RQT,CoePol1982RQT,ExnerPRD1983,Khalfin1997,StefBook2006}, to name a few.

A fundamental subject in the description of the relativistic decays of moving unstable particles is the way the decay laws transform by changing the reference frame. Naturally, it is essential to understand how the decay laws, holding in the rest reference frame of the moving particle, are transformed in the laboratory frame of an observer. The nondecay or survival probability of an unstable particle has been evaluated in \cite{HEP_Stef1996,HEP_Shir2004} for non-vanishing and vanishing values of the linear momentum in terms of the mass distribution density (MDD). The condition of vanishing linear momentum, $p=0$, provides the survival probability in the reference frame where the particle is at rest, while the condition of non-vanishing linear momentum, $p>0$, can be referred to the laboratory frame of an observer where the particles moves with linear momentum $p$. The effects of the relativistic time dilation in quantum decay laws of moving unstable particles remains a matter of central interest. See \cite{TD_FS1963,HEP_Stef1996,HEP_Shir2004,TD_MuonsNat,TD_Farley,HEP_Shir2006,HEP_Shir2009,UrbPLB2014,TD_Giacosa2015,UrbAPB2017}, to name a few.

As a continuation of the scenario described above, here, we evaluate the survival probability, the instantaneous energy and decay rate of a moving unstable particle over short and long times for a wide variety of MDDs and for an arbitrary value of the (constant) linear momentum. In light of the short-time transformations of the survival probability according to the relativistic time dilation, we search for further scaling relations in the decay laws which can relate the conditions of non-vanishing and vanishing linear momentum. In this way, we aim to find further descriptions of the ways the decay laws of moving unstable particle transform by changing reference frame.

The paper is organized as follows. Section 2\ref{2} is devoted to the relativistic quantum laws of a general moving unstable particle. In Section 3\ref{3}, the survival probability is evaluated over short and long times for an arbitrary value of the linear momentum and a general class of MDD. In Section 4 \ref{4}, the transformation of the long-time survival probability is described via a scaling law. Section 5\ref{5} is devoted to the evaluation of the instantaneous mass and instantaneous decay rate over short and long times. In Section 6\ref{6}, the scaling law, describing the transformation of the survival probability, is interpreted in terms of the relativistic time dilation and of the instantaneous mass of the moving unstable particle. Summary and conclusions are reported in Section 7\ref{7}.

\section{Relativistic quantum decay laws}\label{2}

An extended and detailed description of the relativistic quantum decay laws of moving unstable particles has been recently provided in \cite{UrbAPB2017,UrbPLB2014}. Following these references, a brief summary of unstable quantum states, of the survival probability, of the instantaneous energy and decay rate is reported below for the sake of clarity, by adopting the system of units where $\hbar=c=1$. The motion is assumed to be one-dimensional, due to the conservation of the linear momentum \cite{HEP_Stef1996,Khalfin1997,HEP_Shir2004,TD_Giacosa2015,UrbAPB2017,UrbPLB2014}.

In the Hilbert space $\mathcal{H}$ of the quantum states which describe the unstable particle, let the state kets $|m,p\rangle$ be the common eigenstates of the linear momentum $P$ operator and the Hamiltonian $H$ self-adjoint operator, $P|m,p\rangle =p |m,p\rangle$ and $H|m,p\rangle =E(m,p) |m,p\rangle$, for every value of the mass parameter $m$ and of the linear momentum $p$. 
The mass parameter $m$ belongs to the spectrum of the Hamiltonian which is supposed to be continuous with lower bound $\mu_0$, which means $m\geq\mu_0$. In the rest reference frame of the moving particle the linear momentum vanishes and the mentioned state kets become $|m,0\rangle$, while the eigenstates of the Hamiltonian are $E(m,0)=m$. Let $|\phi\rangle$ be the ket of the Hilbert space $\mathcal{H}$ which describes the quantum state of the unstable particle. Such state can be expressed in terms of the eigenstates $|m,0\rangle$ of the Hamiltonian as $|\phi\rangle=\int_{\mu_0}^{\infty} f\left(m\right) |m,0\rangle dm$, via the expansion function $f(m)$. The survival amplitude $A_0(t)$ is defined in the rest reference frame of the moving unstable particle as $A_0(t)=\langle \phi| e^{-\imath H t} |\phi\rangle$ and is given by the integral expression
\begin{eqnarray}
A_0(t)=\int_{\mu_0}^{\infty} \omega\left(m\right) e^{-\imath m t} dm, \label{A0Int}
\end{eqnarray}
where $\imath$ is the imaginary unit. The function $\omega\left(m\right)$ represents the MDD and reads $\omega\left(m\right)=\left|f\left(m\right)\right|^2$. The probability $\mathcal{P}_0(t)$ that the decaying particle is in the initial state at the time $t$, i.e., the survival probability, is given by the following form, $\mathcal{P}_0(t)=\left|A_0(t)\right|^2$, in the rest reference frame of the moving unstable particle.

Let $\Lambda$ be the Lorentz transformation which relates the reference frame where the unstable moving particle is at rest, to the one with velocity $v= p/(m \gamma_L)$, where $\gamma_L$ is the corresponding relativistic Lorentz factor. Let $U\left(\Lambda\right)$ be an unitary representation of the transformation $\Lambda$ acting on the Hilbert space $\mathcal{H}$ and such that $|m,p\rangle=U\left(\Lambda\right)|m,0\rangle$ for every value of the mass parameter $m$ in the Hamiltonian spectrum. The state ket $|\phi,p\rangle$, describes the moving unstable particle with non-vanishing linear momentum $p$. Such state is related to the state ket $|\phi\rangle$ as follows, $|\phi,p\rangle=U\left(\Lambda\right)|\phi\rangle$. The form $E(m,p)=m \gamma_L=\sqrt{p^2+ m^2}$ is obtained by considering the energy-momentum $4$-vector and the Lorentz invariance \cite{GibPolBook,StefBook2006}. The quantity $A_p(t)$ is defined as $A_p(t)=\langle p, \phi| e^{-\imath H t} |\phi,p\rangle$ and represents the survival amplitude in the reference frame where the particle has linear momentum $p$. The approach described above leads to the following integral expression of the survival amplitude,
\begin{eqnarray}
A_p(t)=\int_{\mu_0}^{\infty} \omega\left(m \right)
e^{-\imath \sqrt{p^2+ m^2}t} d m.
\label{Aptdef}
\end{eqnarray}
The quantity $\mathcal{P}_p(t)$ represents the survival probability that the decaying particle is in the initial state at the time $t$ in the reference frame where the unstable particle has linear momentum $p$, and is given by the square modulus of the survival amplitude, $\mathcal{P}_p(t)=\left|A_p(t)\right|^2$. See \cite{HEP_Stef1996,HEP_Shir2004,UrbAPB2017} for details.

The decay laws of the unstable particles are obtained from the MDD via Eqs. (\ref{A0Int}) and (\ref{Aptdef}). In literature, the MDD is usually represented via the Breit-Wigner function \cite{BWMDD3},
\begin{eqnarray}
\omega_{BW}\left(m\right)=\frac{\Theta\left(m-\mu_0\right) \lambda_{BW} \bar{\Gamma}/\left(2 \pi\right)}{\left(m-m_0\right)^2+ \bar{\Gamma}^2/4},
\label{BWMDD}
\end{eqnarray}
where $\lambda_{BW}$ is a normalization factor, $\Theta\left(m\right)$ is the Heaviside unit step function, $m_0$ is the rest mass of the particle and $\bar{\Gamma}$ is the decay rate at rest. A detailed analysis of the survival amplitude of a moving unstable particle has been performed in \cite{HEP_Shir2004} by considering the Breit-Wigner form of the MDD \cite{HEP_Stef1996}. The long-time behavior of the survival amplitude results in dominant inverse power laws, besides additional decaying exponential terms. Refer to \cite{HEP_Shir2004} for details. In \cite{UrbAPB2017,UrbanowskiEPJD2009,UrbanowskiCEJP2009} general forms of MDD are considered with power-law behaviors near the lower bound of the mass spectrum. The long-time decay of the survival amplitude $A_0(t)$ is described by inverse power laws which are determined by the low-mass profile of the MDD \cite{UrbanowskiEPJD2009,UrbanowskiCEJP2009}. Additional removable logarithmic singularities in the low-mass form of the MDD lead to logarithmic-like relaxations of the survival amplitude at rest $A_0(t)$ which can be arbitrarily slower or faster than inverse power laws \cite{GEPJD2015,GEPJD2016}.

Since the unstable particle decays, the initial state is not an eigenstate of the Hamiltonian and the instantaneous mass (energy) is not defined during the time evolution. For this reason, instantaneous mass (energy) and decay rate are defined in the rest reference frame of the moving particle in terms of an effective Hamiltonian. Such operator acts on the subspace of the Hilbert space $\mathcal{H}$ which is spanned by the initial state. Refer to \cite{UrbPLB2014,UrbAPB2017,UrbanowskiEPJD2009,UrbanowskiCEJP2009,UrbanowskiPRA1994} for details. In the same way, the instantaneous mass (energy) and decay rate $\Gamma_p(t)$ of the moving unstable particle are defined in the frame system where the particle has linear momentum $p$. Both for vanishing and non-vanishing values of the linear momentum $p$, the instantaneous mass $M_{p}(t)$ and decay rate $\Gamma_p(t)$ are obtained from the survival amplitude $A_p(t)$ via the following forms \cite{UrbPLB2014,UrbAPB2017,UrbanowskiEPJD2009,UrbanowskiCEJP2009,GEPJD2015,GEPJD2016},
\begin{eqnarray}
&&M_{p}(t)=-\mathrm{Im} \left\{\frac{\partial_t A_p(t)}{A_p(t)}\right\} \label{Epinst}\\
&&\Gamma_{p}(t)=-2  \mathrm{Re}\left\{\frac{\partial_t A_p(t)}{A_p(t)}\right\}. \hspace{0.2em}
  \,\,\,\label{gEdef}
\end{eqnarray}
The long-time behavior of the instantaneous mass $M_{p}(t)$ and decay rate $\Gamma_{p}(t)$ has been evaluated in \cite{UrbPLB2014} and \cite{TD_Giacosa2015} for the Breit-Wigner form of the MDD.

\subsection{Relativistic time dilation}\label{21}

The interpretation of decay processes via the theory of special relativity suggests that the lifetime which is detected in the rest reference frame of the unstable particle increases in the reference frame where the particle is moving. The increase is due to the relativistic dilation of times and is determined by the relativistic Lorentz factor \cite{Moller1972}.

The appearance of the relativistic time dilation in quantum decays processes is a matter of great interest and fruitful discussions. See \cite{TD_FS1963,HEP_Stef1996,HEP_Shir2004,TD_MuonsNat,TD_Farley,HEP_Shir2006,HEP_Shir2009,UrbPLB2014,TD_Giacosa2015,
UrbAPB2017}, to name a few. A broadly shared opinion is that the relativistic time dilation influences the survival probability uniquely in the short-time exponential decay. Briefly, this means that the survival probability $\mathcal{P}_p\left(t\right)$ and the survival probability at rest $\mathcal{P}_0\left(t\right)$ are given by the following exponential forms, $\mathcal{P}_p\left(t\right)\sim e^{-t/\tau_p}$, and $\mathcal{P}_0\left(t\right)\sim e^{-t/\tau_0}$, over short times. Under this assumption, the survival probability obeys the scaling law 
\begin{eqnarray}
\mathcal{P}_p\left(t\right)\sim \mathcal{P}_0\left(\frac{t}{\gamma_L}\right), \label{ED}
\end{eqnarray}
over short times. The parameter $\tau_0$ represents the lifetime of the particle at rest, while $\tau_p$ is the lifetime which is detected in the reference frame where the particle is moving with linear momentum $p$. According to the relativistic time dilation, the lifetimes are related as follows, $\tau_p= \tau_0 \gamma_L$, where $\gamma_L$ is the corresponding relativistic Lorentz factor. Refer to \cite{HEP_Stef1996,HEP_Shir2004,HEP_Shir2006,HEP_Shir2009,UrbPLB2014,TD_Giacosa2015,UrbAPB2017} for an extended explanation. In this context, we intend to study the survival probability $\mathcal{P}_p\left(t\right) $ and the survival probability at rest $\mathcal{P}_0\left(t\right)$ over short and long times for a wide variety of MDDs. We search for further ways to describe the transformations of the decay laws which occur by changing reference frame.

\section{Survival probability versus linear momentum}\label{3}

In the present Section the short- and long-time behaviors of the survival probability are studied for a general value $p$ of the 
constant linear momentum of the moving unstable particle. The analysis performed in the whole paper is based entirely on the form (\ref{Aptdef}) of the survival amplitude \cite{HEP_Stef1996,HEP_Shir2006}. For the sake of convenience, the survival amplitude is expressed via the dimensionless variables $\tau$, $\xi$, $\rho$ and $\eta$. These variables are defined in terms of a generic scale mass $m_s$ as follows, $\tau=m_s t$, $\xi=m/m_s$, $\rho=p/m_s$ and $\eta=\sqrt{\rho^2+\xi^2}$. The MDD is expressed in terms of the auxiliary function $\Omega\left(\xi\right)$ via the scaling law $\omega\left(m_s \xi \right)=\Omega\left(\xi\right)/m_s $, for every $\xi\geq \xi_0$, where $\xi_0=\mu_0/m_s$. In this way, the survival amplitude $A_p(t)$ results in the expression below,
\begin{eqnarray}
A_p(t)=\int_{\xi_0}^{\infty} \Omega\left(\xi \right)
e^{-\imath \eta \tau} d \xi.
\label{Aptxi}
\end{eqnarray}

The MDDs under study are defined over the infinite support $\big[\mu_0,\infty\big)$ by auxiliary functions $\Omega\left(\xi\right)$ of the following form,
\begin{eqnarray}
\Omega\left(\xi \right)= \left(\xi-\xi_0\right)^{\alpha}
\Omega_0\left(\xi \right).
\label{Omegaalpha}
\end{eqnarray}
In order to study the long-time behavior of the decay laws of the moving unstable particle, the MDDs are requested to obey the conditions below. The lower bound of the mass spectrum is chosen to be non-vanishing, $\xi_0>0$. The constraints $\alpha\geq 0$, and $\Omega_0  \left(\xi_0 \right)>0$, are also requested. The function $\Omega_0\left(\xi \right)$ and the derivatives $\Omega^{(j)}_0\left(\xi \right)$ are required to be summable, for every $j=1, \ldots, \lfloor \alpha \rfloor +4$, and continuously differentiable in the whole support $\big[\mu_0,\infty\big)$, for every $j=1, \ldots, \lfloor \alpha \rfloor +3$. Consequently, the following limits, $\lim_{\xi\to\xi_0^+}\Omega^{(j)}_0\left(\xi \right)$, must exist finite and be $\Omega^{(j)}_0\left(\xi_0 \right)$ for every $j=0, \ldots, \lfloor \alpha \rfloor+4$. The functions $\Omega_0^{(j)}\left(\xi \right)$ have to decay sufficiently fast as $\xi\to+\infty$, so that the auxiliary function $\Omega\left(\xi\right)$ and the derivatives $\Omega^{(j)}\left(\xi\right)$ vanish as $\xi\to+\infty$, for every $j=0, \ldots, \lfloor \alpha \rfloor$.

As far as the short-time behavior of the survival amplitude is concerned, let the auxiliary function 
decay as follows, $\Omega\left(\xi \right)= \mathcal{O}\left(\xi^{-1-l_0}\right)$ for $\xi\to+\infty$, with $l_0>5$. Under this condition, the survival amplitude evolves algebraically over short times,
\begin{eqnarray}
A_p(t)\sim 1-\imath a_0 t-a_1 t^2+ \imath a_2 t^3, \label{Aptshort}
\end{eqnarray}
for $t \ll 1/ m_s$. The constants $a_0$, $a_1$ and $a_2$ are given by the expressions below,
\begin{eqnarray}
&&\hspace{-3em}a_0=\int_{\mu_0}^{\infty}\omega\left(m\right)\sqrt{p^2+m^2}  dm,
\nonumber \\
&&\hspace{-3em}a_1=\frac{1}{2}\int_{\mu_0}^{\infty}\omega\left(m\right) \left(p^2+m^2\right)  dm,
\nonumber \\ 
&&\hspace{-3em}a_2=\frac{1}{6}\int_{\mu_0}^{\infty}\omega\left(m\right)\left(p^2+m^2\right)^{3/2}  dm.
\nonumber 
\end{eqnarray}
The short-time evolution of the survival probability is derived from Eqs. (\ref{Aptxi}) and (\ref{Aptshort}), and is algebraic, 
\begin{eqnarray}
\mathcal{P}_p(t)\sim 1 - \pi_0 t^2, \label{Pptshort}
\end{eqnarray}
for $t \ll 1/m_s$, where $\pi_0=2 a_1-a_0^2$.

The long-time behavior of the survival amplitude is obtained from Eq. (\ref{Aptxi}) and from the following equivalent form, 
\begin{eqnarray}
A_p(t)=\int_{\eta_0}^{\infty} \frac{\eta \Omega\left(\sqrt{\eta^2-\rho^2} \right)}
{\sqrt{\eta^2-\rho^2}} e^{-\imath \eta \tau}
 d \eta,
\label{Apteta}
\end{eqnarray}
where $\eta_0=\sqrt{\rho^2+\xi_0^2}$. Notice that the constraint of non-vanishing lower bound of the mass spectrum, $\xi_0>0$, is fundamental for the equivalence of the expressions (\ref{Aptxi}) and (\ref{Apteta}) of the survival amplitude. The asymptotic analysis \cite{ErdelyiBook1956,WongBook1989} of the integral form appearing in Eq. (\ref{Apteta}) for $\tau \gg 1$, 
provides the expression of the survival amplitude over long times, 
\begin{eqnarray}
A_p(t)\sim c_0  e^{-\imath \left(\frac{\pi}{2}
\left(1+\alpha\right)+\sqrt{\mu_0^2+p^2} t\right)}\left(\frac{\chi_p}{m_s t}\right)^{1+\alpha},
\label{Aplongt}
\end{eqnarray}
for $t \gg 1/ m_s$, where $c_0=\Gamma\left(1+\alpha\right)
\Omega_0\left(\xi_0\right)$, and
\begin{eqnarray}
\chi_p = \sqrt{1+\frac{p^2}{\mu_0^2}}. \label{Chip}
\end{eqnarray}
The asymptotic form (\ref{Aplongt}) of the survival 
amplitude holds for every value of the linear momentum $p$, non-vanishing, arbitrarily large or small, or vanishing; for the variety of MDDs under study; for every $\alpha \geq 0$ and for every value $\mu_0$ of the lower bound of the mass spectrum such that $\mu_0/m_s>0$. The last condition is crucial and will be interpreted in the next Section in terms of the instantaneous mass at rest of the moving unstable particle. The square modulus of the asymptotic expression (\ref{Aplongt}) approximates the survival probability over long times,
\begin{eqnarray}
\mathcal{P}_p(t)\sim c^2_0 \left(\frac{\chi_p}{m_s t}\right)^{2\left( 1+\alpha\right)},
\label{Pplongt}
\end{eqnarray}
for $t \gg 1/ m_s$. Notice that the time scale $1/m_s$, and, consequently, the short or long times, $t\ll 1/m_s$, or $t\gg 1/m_s$, are independent of the auxiliary function $\Omega\left(\xi\right)$ and are determined uniquely by the MDD. 

\begin{figure}[!b]
  \begin{center}
    \includegraphics[width=3.5in]{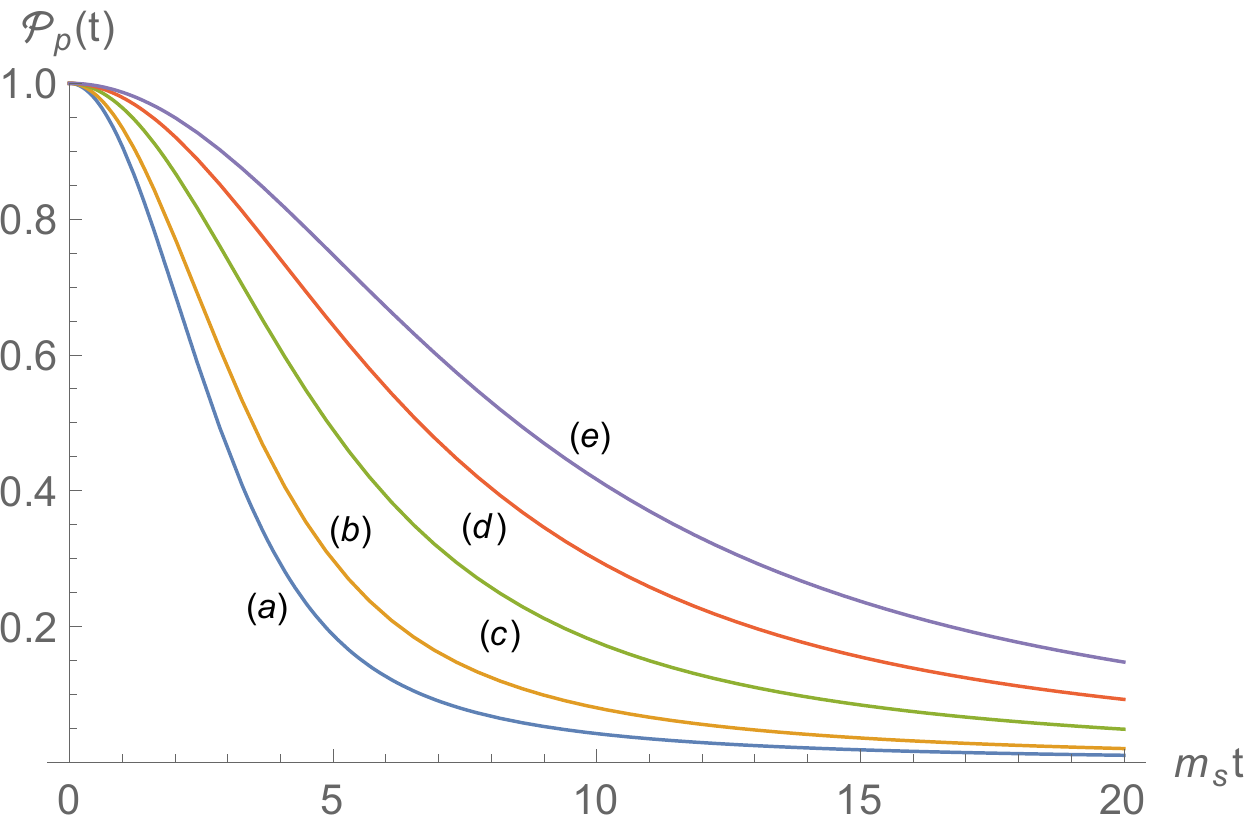}
  \end{center}
\caption{ \small (Color online) The survival probability $\mathcal{P}_p(t)$ versus $\left(m_s t\right)$ for $0\leq m_s t \leq 20$, MDDs given by Eq. (\ref{Ofigs}), $\alpha=0$, $\mu_0=m_s$, and different values of the linear momentum $p$. Curve $(a)$
 corresponds to $p=0 m_s$; $(b)$ corresponds to $p=m_s$; 
$(c)$ corresponds to $p= 2 m_s$; 
$(d)$ corresponds to $p= 3 m_s$; 
$(e)$ corresponds to $p = 4 m_s$.}
\label{Fig1}
\end{figure}

\begin{figure}[!b]
  \begin{center}
    \includegraphics[width=3.5in]{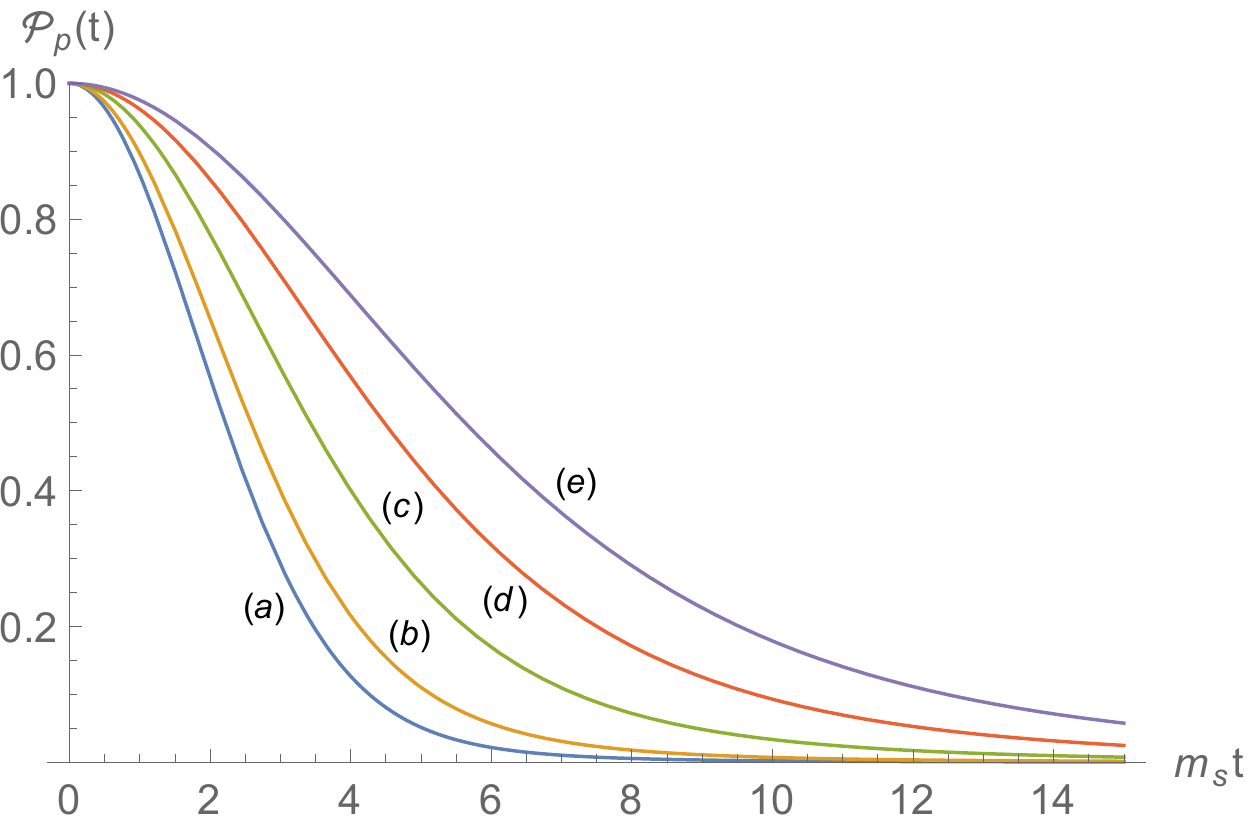}
  \end{center}
\caption{ \small (Color online) The survival probability $\mathcal{P}_p(t)$ versus $\left(m_s t\right)$ for $0\leq m_s t \leq 15$, MDDs given by Eq. (\ref{Ofigs}), $\alpha=1$, $\mu_0=m_s$, and different values of the linear momentum $p$. Curve $(a)$
 corresponds to $p=0 m_s$; $(b)$ corresponds to $p=m_s$; 
$(c)$ corresponds to $p= 2 m_s$; 
$(d)$ corresponds to $p= 3 m_s$; 
$(e)$ corresponds to $p = 4 m_s$.}
\label{Fig2}
\end{figure}

\begin{figure}[!b]
  \begin{center}
    \includegraphics[width=3.5in]{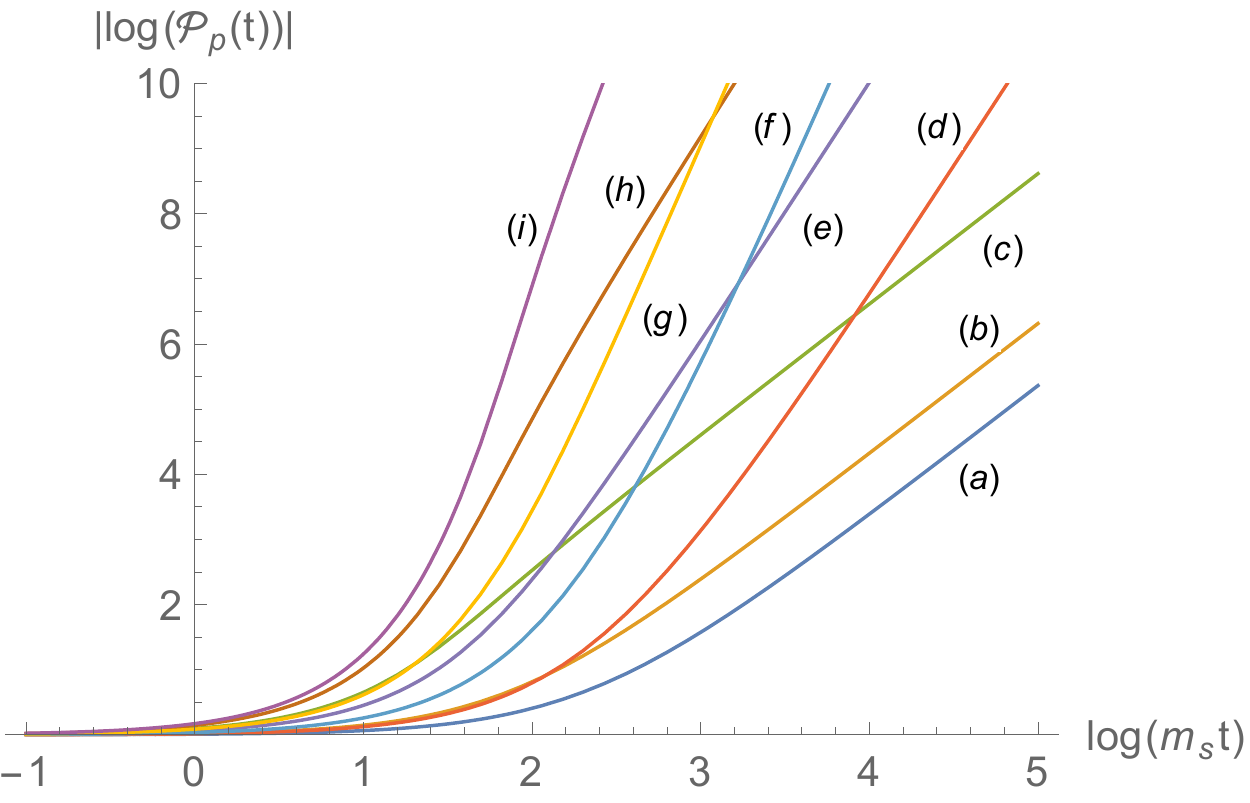}
  \end{center}
\caption{ \small (Color online) Quantity $\left|\log\left(\mathcal{P}_p(t)\right)\right|$ versus $\log\left(m_s t\right)$ for $e^{-1}\leq m_s t \leq e^{5}$, MDDs given by Eq. (\ref{Ofigs}), $\mu_0=m_s$ and different values of the parameter $\alpha$ and of the linear momentum $p$. Curve $(a)$
 corresponds to $\alpha=0$ and $p= 5 m_s$; $(b)$ corresponds to $\alpha=0$ and $p=3 m_s$; $(c)$ corresponds to $\alpha=0$ and $p=0$; $(d)$ corresponds to $\alpha=1$ and $p=5 m_s$; $(e)$ corresponds to $\alpha=1$ and $p=2 m_s$; $(f)$ corresponds to $\alpha=2$ and $p= 4 m_s$; $(g)$ corresponds to $\alpha=2$ and $p=2m_s$; $(h)$ corresponds to $\alpha=1$ and $p=0 m_s$; $(i)$ corresponds to $\alpha=2$ and $p=0$.}
\label{Fig3}
\end{figure}
\begin{figure}[!b]
  \begin{center}
    \includegraphics[width=3.5in]{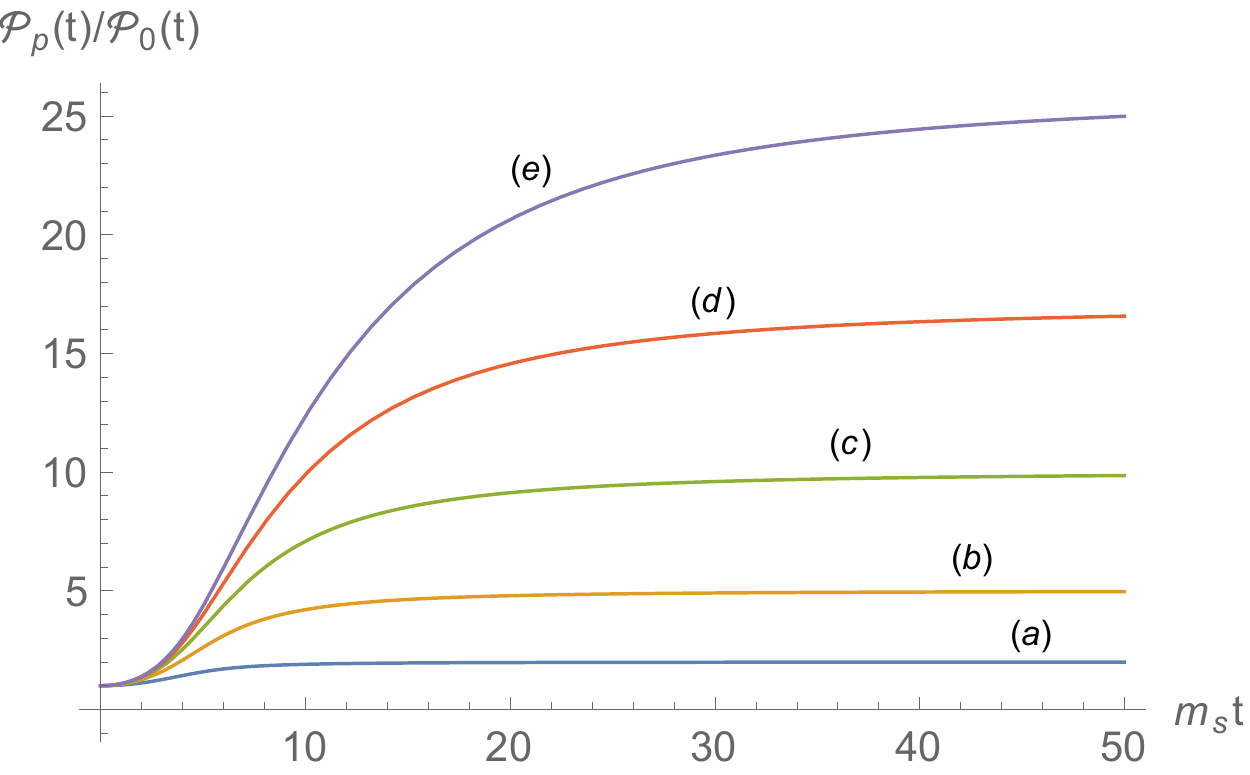}
  \end{center}
\caption{ \small (Color online)  Ratio $\mathcal{P}_p(t)/\mathcal{P}_0(t)$ versus $\left(m_s t\right)$ for $0\leq m_s t \leq 50$, 
MDDs given by Eq. (\ref{Ofigs}), $\mu_0=m_s$, $\alpha=0$ and different values of the linear momentum $p$. Curve $(a)$ corresponds to $p= m_s$; $(b)$ corresponds to $p= 2 m_s$; $(c)$ corresponds to $p=3 m_s$; $(d)$ corresponds to $p= 4 m_s$; $(e)$ corresponds to $p=5 m_s$.}
\label{Fig4}
\end{figure}
\begin{figure}[!b]
  \begin{center}
    \includegraphics[width=3.5in]{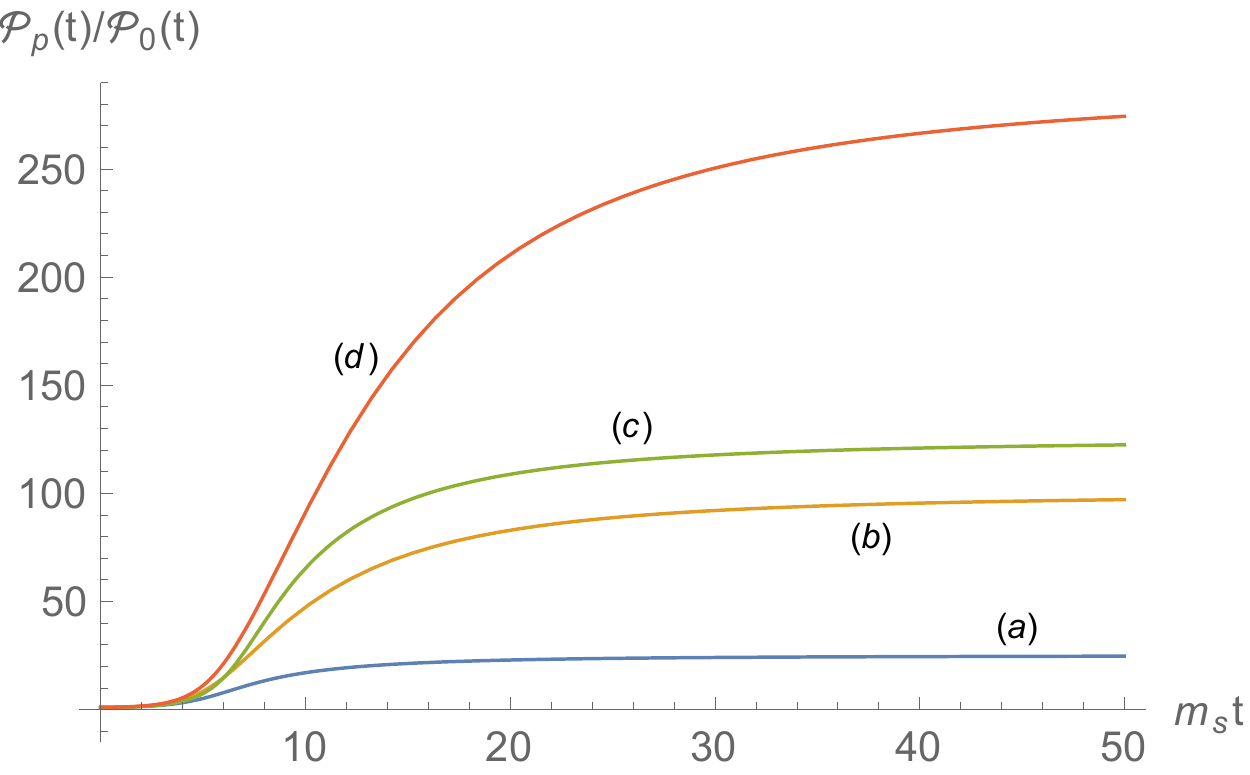}
  \end{center}
\caption{ \small (Color online)  Ratio $\mathcal{P}_p(t)/\mathcal{P}_0(t)$ versus $\left(m_s t\right)$ for $0\leq m_s t \leq 50$, 
MDDs given by Eq. (\ref{Ofigs}), $\mu_0=m_s$, and different values of the parameter $\alpha$ and of the linear momentum $p$. Curve $(a)$ corresponds to $\alpha=1$ and $p=2 m_s$; $(b)$ corresponds to $\alpha=1$ and $p= 3 m_s$; $(c)$ corresponds to $\alpha=2$ and $p=2 m_s$; $(d)$ corresponds to $\alpha=1$ and $p=4m_s$.}
\label{Fig5}
\end{figure}
\begin{figure}[!b]
  \begin{center}
    \includegraphics[width=3.5in]{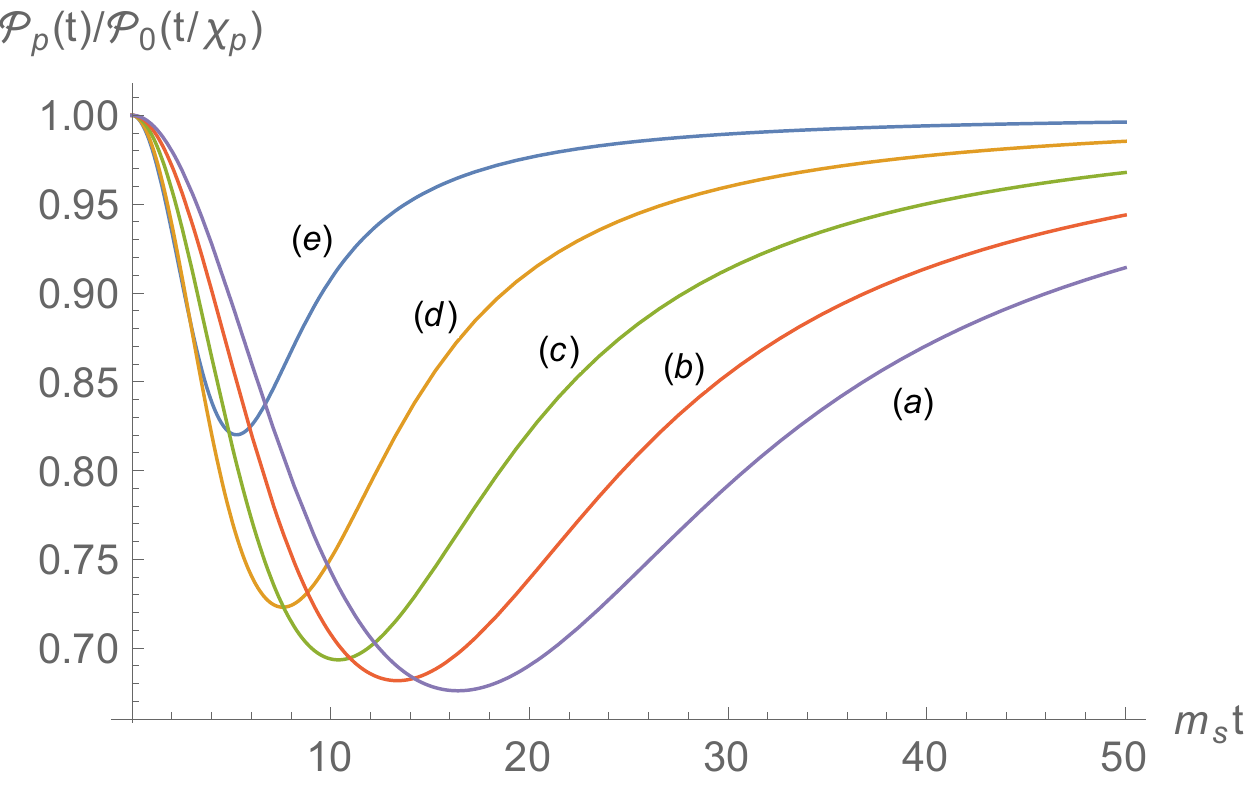}
  \end{center}
\caption{ \small (Color online) Ratio $\mathcal{P}_p\left(t\right)/\mathcal{P}_0\left(t/\chi_p\right)$ versus $\left(m_s t\right)$ for $0\leq m_s t \leq 50$, MDDs given by Eq. (\ref{Ofigs}), $\mu_0=m_s$, $\alpha=0$ and different values of the linear momentum $p$. Curve $(a)$ corresponds to $p= 5 m_s$; $(b)$ corresponds to $p= 4 m_s$; $(c)$ corresponds to $p=3 m_s$; $(d)$ corresponds to $p= 2 m_s$; $(e)$ corresponds to $p= m_s$.}
\label{Fig6}
\end{figure}
\begin{figure}[!b]
  \begin{center}
    \includegraphics[width=3.5in]{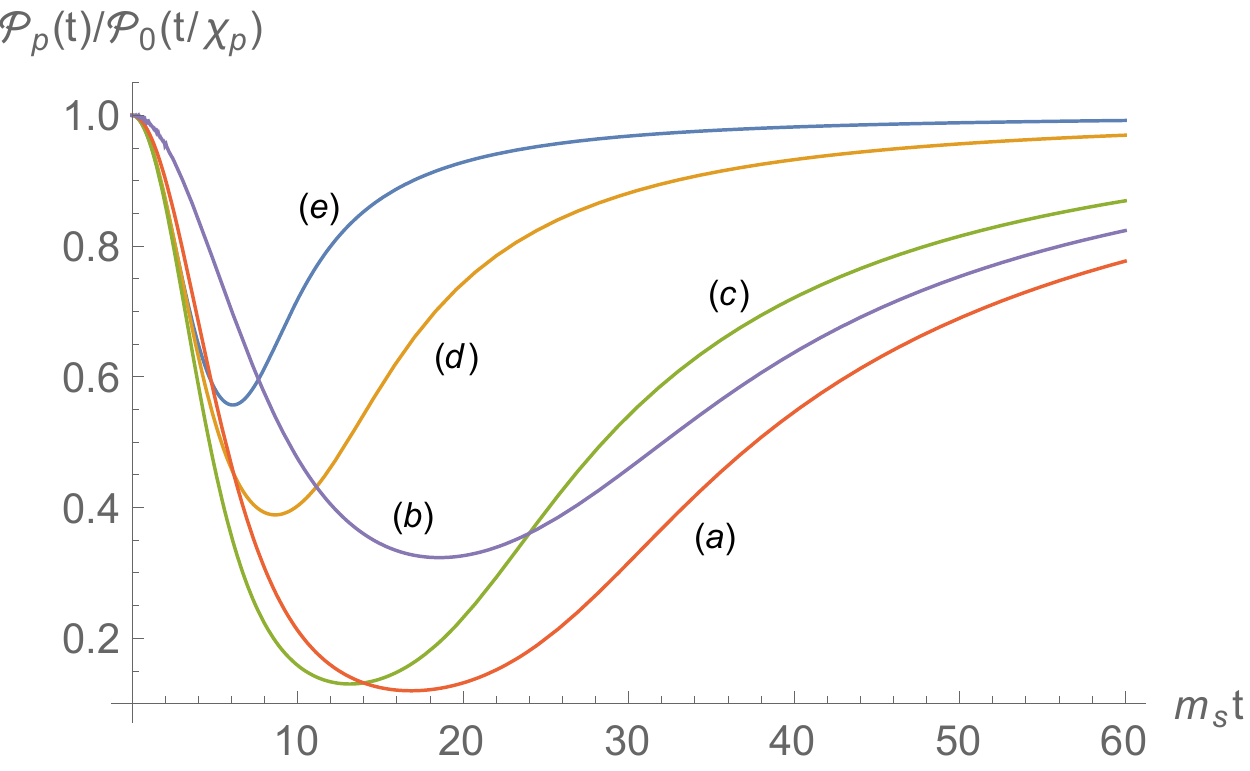}
  \end{center}
\caption{ \small (Color online) Ratio $\mathcal{P}_p\left(t\right)/\mathcal{P}_0\left(t/\chi_p\right)$ for $0\leq m_s t \leq 60$, 
MDDs given by Eq. (\ref{Ofigs}), $\mu_0=m_s$, and different values of the parameter $\alpha$ and of the linear momentum $p$. Curve $(a)$ corresponds to $\alpha=2$ and $p=4 m_s$; $(b)$ corresponds to $\alpha=1$ and $p= 5 m_s$; $(c)$ corresponds to $\alpha=2$ and $p=3 m_s$; $(d)$ corresponds to $\alpha=1$ and $p=2 m_s$, $(e)$ corresponds to $\alpha=1$ and $p= m_s$.}
\label{Fig7}
\end{figure}

\section{Scaling law for the survival probability
}\label{4}

The expression (\ref{Pplongt}) of the survival probability holds for every value of the constant linear momentum $p$. Such arbitrariness allows to evaluate the survival probability in the non-relativistic and ultrarelativistic limits. For vanishing value of the linear momentum, $p=0$, or, equivalently, in the rest reference frame of the unstable particle, the survival probability is approximated over long times as follows,
\begin{eqnarray}
\mathcal{P}_0(t)\sim \frac{c^2_0}{\left(m_s t\right)^{2\left( 1+\alpha\right)}},
\label{P0longtU}
\end{eqnarray}
for $t \gg 1/ m_s$. Instead, consider large values of the linear momentum, $ p \gg \mu_0$. In such condition the survival probability is approximated over long times as below,
\begin{eqnarray}
\mathcal{P}_p(t)\sim c^2_0 \left(\frac{p}{\mu_0 m_s t}\right)^{2\left(1+\alpha\right)},
\label{PplongtU}
\end{eqnarray}
for $t \gg 1/ m_s$. 
By comparing Eqs. (\ref{Pplongt}) and (\ref{P0longtU}), we observe that, for the MDDs under study, the survival probability 
obeys, approximately over long times, the following scaling law, 
\begin{eqnarray}
\mathcal{P}_p\left(t\right) \sim \mathcal{P}_0\left(\frac{t}{ \chi_p}\right), \label{PpP0L}
\end{eqnarray}
for $t \gg 1/m_s$. This is the main result of the paper. In fact, the above scaling law describes how the survival probability at rest transforms, approximately over long times, in the reference frame where the particles moves with constant linear momentum $p$. The transformation can be interpreted, approximately, as the effect of a time dilation which is determined by the scaling factor $\chi_p$. Notice that the scaling factor, given by Eq. (\ref{Chip}), diverges in the limit $\mu_0\to 0^+$. In Section 6, the scaling law (\ref{PpP0L}) is interpreted, via the special relativity, as the effect of the relativistic time dilation. This interpretation holds if the asymptotic value of the instantaneous mass is considered as the effective mass of the moving unstable particle over long times.

The correction to the scaling law (\ref{PpP0L}) can be estimated via the expression $\left(\mathcal{P}_p\left(t\right)-\mathcal{P}_0\left(t/ \chi_p\right)\right)\big/\mathcal{P}_0\left(t/ \chi_p\right)$. For the MDDs under study, such correction vanishes inversely quadratically over long times, 
\begin{eqnarray}
\frac{\mathcal{P}_p\left(t\right)-\mathcal{P}_0\left(t/ \chi_p\right)}{\mathcal{P}_0\left(t/ \chi_p\right)} \sim  \frac{\kappa_p }{\left(m_s t\right)^{2}},
\label{DeltaPpP0L}
\end{eqnarray}
for $t\gg 1/m_s$, where \begin{eqnarray}
&&\kappa_p=\frac{\left(1+\alpha\right)\left(2+\alpha\right)m_s p^2}{\mu_0^3}\Bigg
(
2\frac{\Omega^{\prime}_0\left(\mu_0/m_s\right)}{\Omega_0\left(\mu_0/m_s\right)}
-\frac{m_s}{\mu_0 \chi_p^2}
 \nonumber\\&& \hspace{2.2em}\times \,\Bigg(3+\alpha
+\left(\frac{5}{2}+\alpha \right)\frac{p^2}{ \mu_0^2}\Bigg)
\Bigg).\nonumber
\end{eqnarray}

Numerical analysis of the survival probability $\mathcal{P}_p(t)$ has been displayed in Figures \ref{Fig1}, \ref{Fig2}, \ref{Fig3}, \ref{Fig4}, \ref{Fig5}. The computed MDDs are given by the following toy form of the auxiliary function,
\begin{eqnarray}
\Omega\left(\xi\right)= w_{\alpha} \xi \left(\xi^2-\xi^2_0\right)^{\alpha} e^{- \xi^2},\label{OmegaFig}
 \label{Ofigs}
\end{eqnarray}
where $w_{\alpha}$ is a normalization factor and reads $w_{\alpha}=2 e^{\xi_0^2}/ \Gamma\left(1+\alpha\right)$. Different values of the non-negative power $\alpha$ and linear momentum $p$ are considered. The corresponding MDDs belong to the class under study which is defined in Section 3\ref{3} via Eq. (\ref{Omegaalpha}). 
The asymptotic lines appearing in Figure \ref{Fig3} agree with the long-time inverse-power-law decays of the survival probability, given by Eq. (\ref{Pplongt}). The ordinates of the asymptotic horizontal lines of Figures \ref{Fig4} and \ref{Fig5} are in accordance with the form (\ref{Chip}) of the scaling factor $\chi_p$. The long-time dilation in the survival probability, given by the scaling law (\ref{PpP0L}), is confirmed by the common horizontal asymptotic line, at ordinate $1$, which appears in Figures \ref{Fig6} and \ref{Fig7}.

\section{Instantaneous mass and decay rate versus linear momentum}\label{5}

In the present Section the instantaneous mass and decay rate are analyzed over short and long times for every value of the constant linear momentum $p$ of the particle, which is detected in the laboratory frame of the observer. The instantaneous mass and decay rate are evaluated from the survival amplitude via Eqs. (\ref{Epinst}) and (\ref{gEdef}).

Again, let the auxiliary function decay as $\Omega\left(\xi \right)= \mathcal{O}\left(\xi^{-1-l_0}\right)$ for $\xi\to+\infty$, with $l_0>5$. The short-time evolution of the instantaneous mass and decay rate are obtained from the behavior (\ref{Aptshort}) of the survival amplitude and from Eqs. (\ref{Epinst}) and (\ref{gEdef}). In this way, the following algebraic evolutions result over short times,
\begin{eqnarray}
&&M_p(t) \sim a_0 - \pi_1 t^2, \label{MptS}
\end{eqnarray}
and 
\begin{eqnarray}
&&\Gamma_p(t) \sim  \pi_2 t, \label{GammaptS} 
\end{eqnarray}
for $t \ll 1/m_s$, where $\pi_1= a_0^3+3 \left(a_2-a_0 a_1\right)$, and $\pi_2= 2 \left(2 a_1 -a_0^2\right)$.

\begin{figure}[!b]
  \begin{center}
    \includegraphics[width=3.5in]{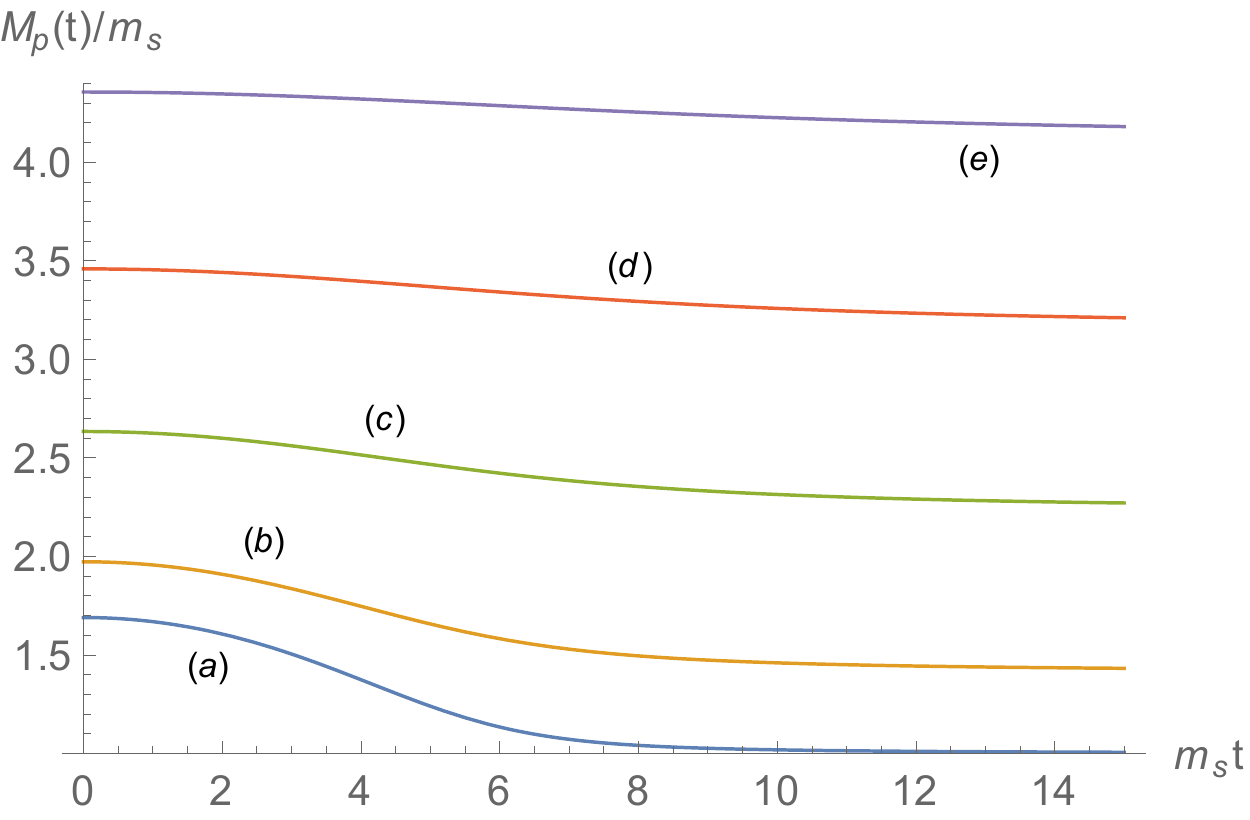}
  \end{center}
\caption{ \small (Color online) Quantity $M_p(t)/m_s$ versus $\left(m_s t\right)$ for $0 \leq m_s t \leq 15$, MDDs given by Eq. (\ref{Ofigs}), $\mu_0=m_s$, $\alpha=1$ and different values of the linear momentum $p$. Curve $(a)$
 corresponds to $p=0 m_s$; $(b)$ corresponds to $p=m_s$; $(c)$ corresponds to corresponds to $p=2 m_s$; $(d)$ corresponds to $p=3 m_s$; $(e)$ corresponds to $p=4 m_s$.}
\label{Fig8}
\end{figure}

\begin{figure}[!b]
  \begin{center}
    \includegraphics[width=3.5in]{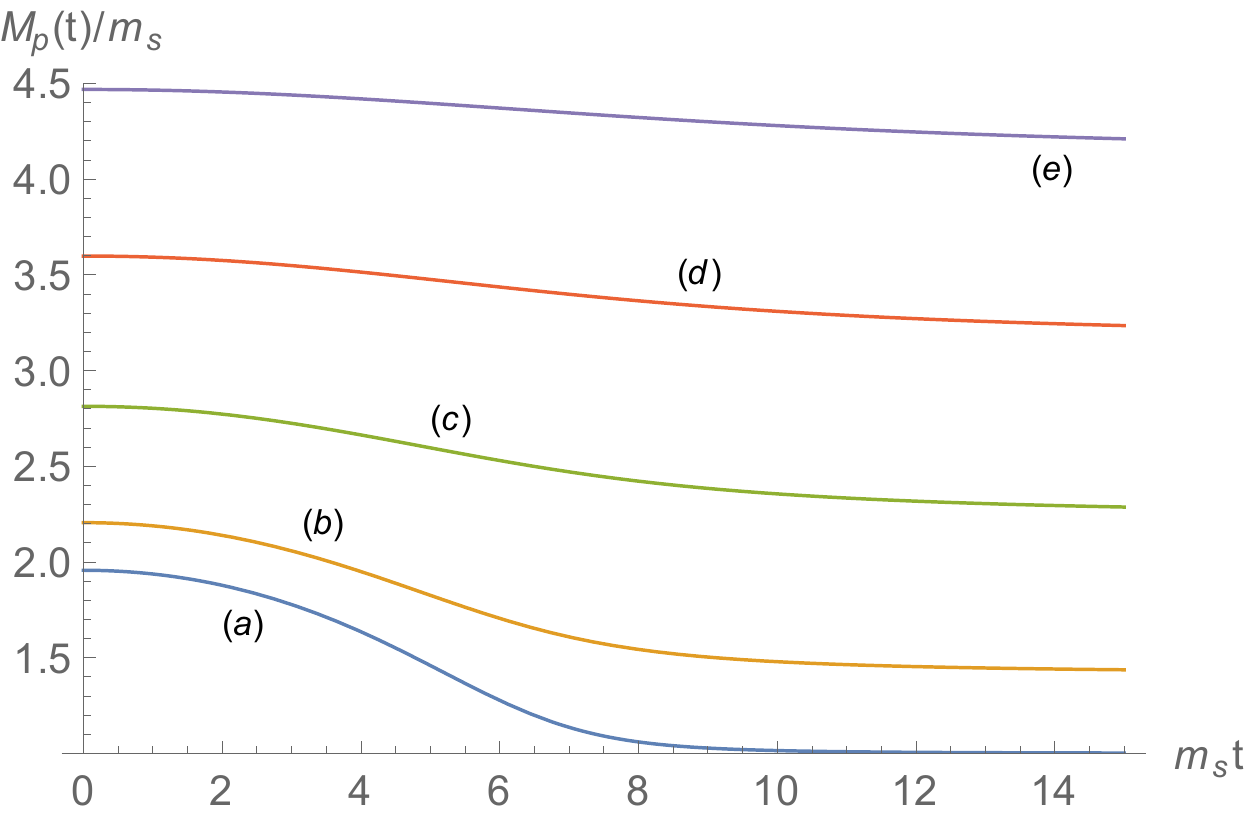}
  \end{center}
\caption{ \small (Color online) Quantity $M_p(t)/m_s$ versus $\left(m_s t\right)$ for $0 \leq m_s t \leq 15$, MDDs given by Eq. (\ref{Ofigs}), $\mu_0=m_s$, $\alpha=2$ and different values of the linear momentum $p$. Curve $(a)$
 corresponds to $p=0 m_s$; $(b)$ corresponds to $p=m_s$; $(c)$ corresponds to  corresponds to $p=2 m_s$; $(d)$ corresponds to $p=3 m_s$; $(e)$ corresponds to $p=4 m_s$.}
\label{Fig9}
\end{figure}

\begin{figure}[!b]
  \begin{center}
    \includegraphics[width=3.5in]{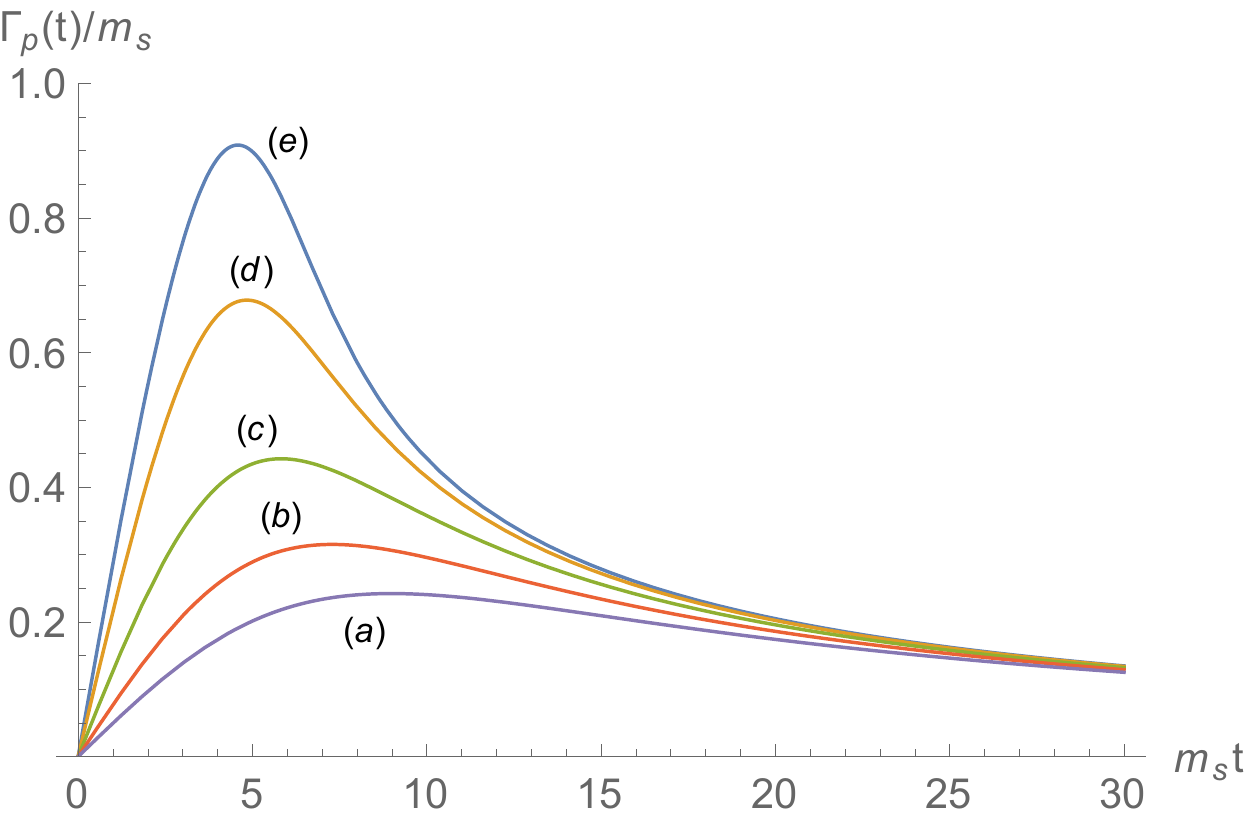}
  \end{center}
\caption{ \small (Color online) Quantity $\Gamma_p(t)/m_s$ versus $\left(m_s t\right)$ for $0 \leq m_s t \leq 30$, 
MDDs given by Eq. (\ref{Ofigs}), $\mu_0=m_s$, $\alpha=1$ and different values of the linear momentum $p$. Curve $(a)$
 corresponds to $p=4 m_s$; $(b)$ corresponds to $p=3m_s$; $(c)$ corresponds to $p=2 m_s$; $(d)$ corresponds to $p=1 m_s$; $(e)$ corresponds to $p=0 m_s$.}
\label{Fig10}
\end{figure}

\begin{figure}[!b]
  \begin{center}
    \includegraphics[width=3.5in]{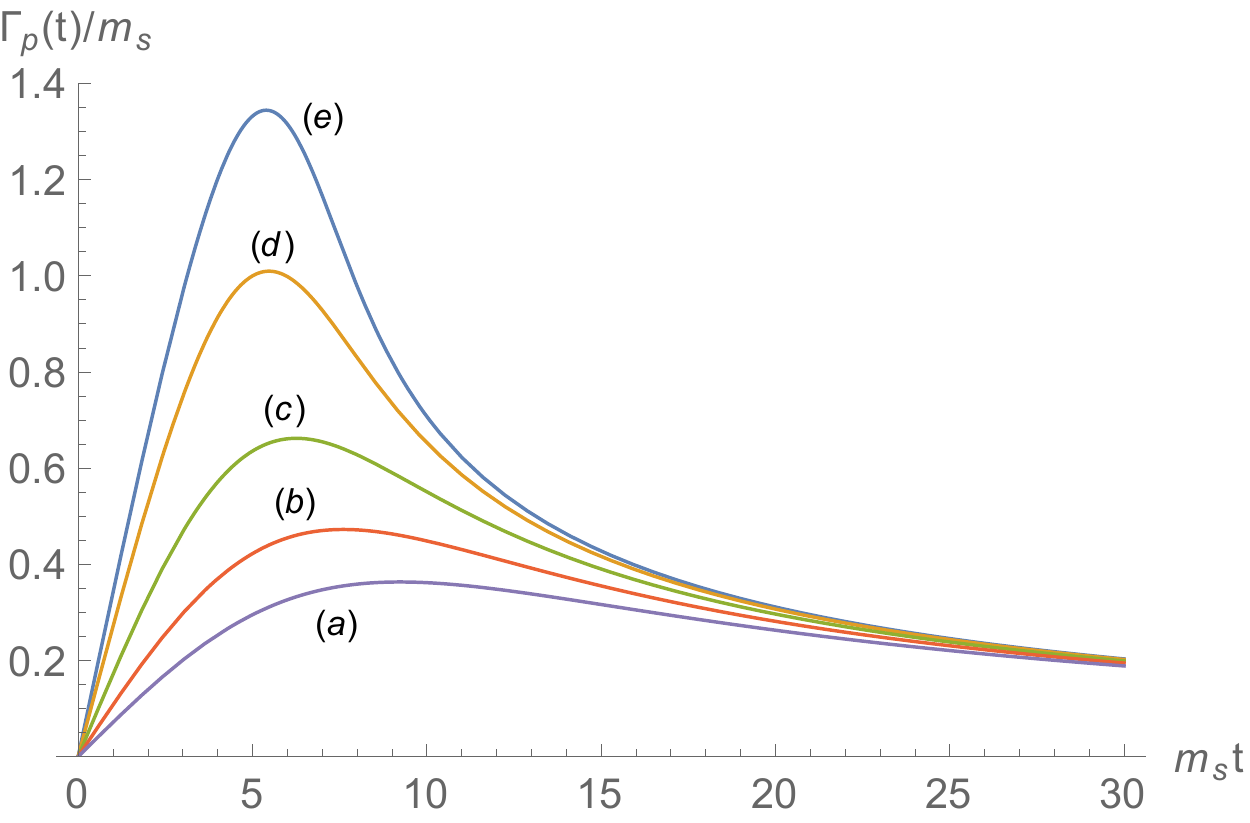}
  \end{center}
\caption{ \small (Color online) Quantity $\Gamma_p(t)/m_s$ versus $\left(m_s t\right)$ for $0 \leq m_s t \leq 30$, MDDs given by Eq. (\ref{Ofigs}), $\mu_0=m_s$, $\alpha=2$ and different values of the linear momentum $p$. Curve $(a)$
 corresponds to $p=4 m_s$; $(b)$ corresponds to $p=3m_s$; $(c)$ corresponds to $p=2 m_s$; $(d)$ corresponds to $p=1 m_s$; $(e)$ corresponds to $p=0 m_s$.}
\label{Fig11}
\end{figure}

\begin{figure}[!b]
  \begin{center}
    \includegraphics[width=3.5in]{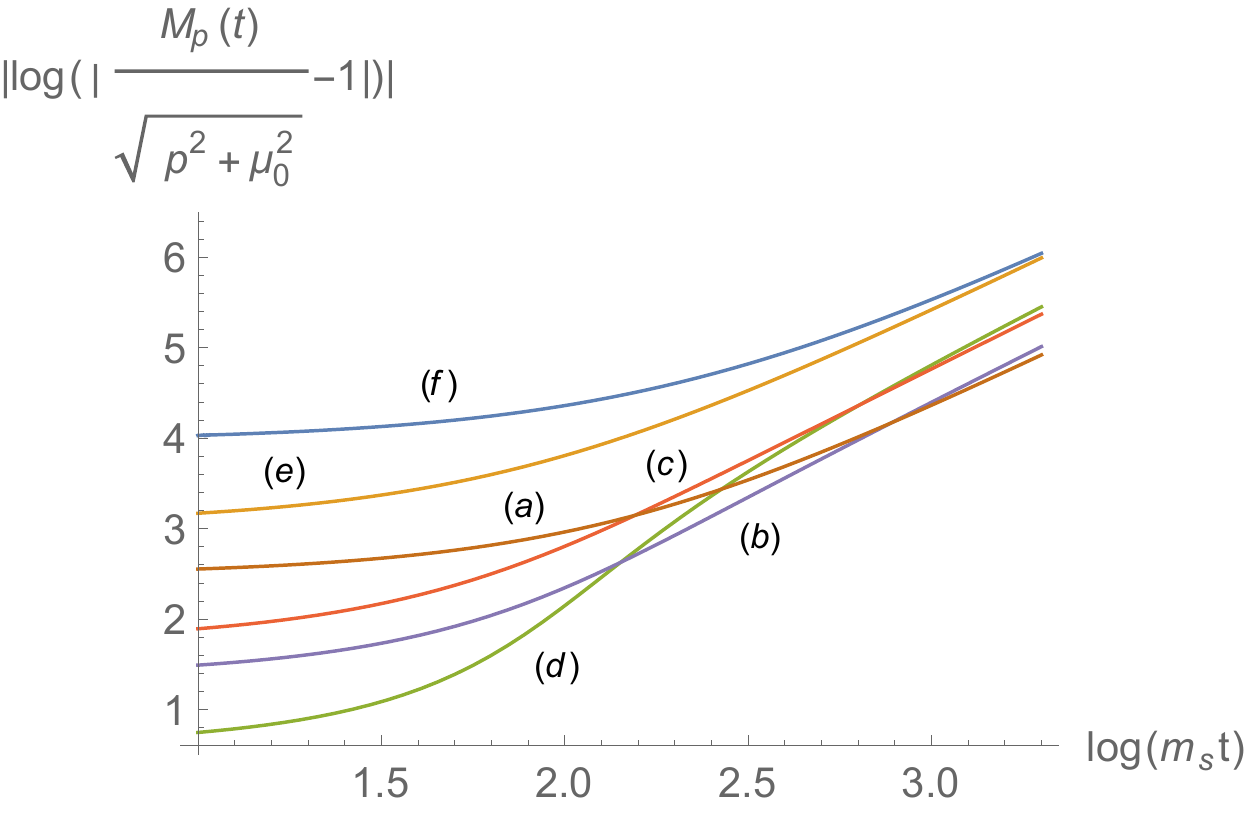}
  \end{center}
\caption{ \small (Color online)  Quantity $\left|\log\left(\left|\frac{M_p(t)}{\sqrt{p^2+\mu_0^2}}-1\right|\right)\right|$ versus $\log\left(m_s t\right)$ for $e\leq m_s t \leq e^{3.3}$, MDDs given by Eq. (\ref{Ofigs}), $\mu_0=m_s$, and different values of the parameter $\alpha$ and of the linear momentum $p$. Curve $(a)$ corresponds to $\alpha=2$ and $p=4 m_s$; $(b)$ corresponds to $\alpha=2$ and $p= 2 m_s$; $(c)$ corresponds to $\alpha=1$ and $p=2 m_s$; $(d)$ corresponds to $\alpha=2$ and $p=m_s$; $(e)$ corresponds to $\alpha=0$ and $p=3 m_s$; $(f)$ corresponds to $\alpha=0$ and $p=5 m_s$.}
\label{Fig12}
\end{figure}

\begin{figure}[!b]
  \begin{center}
    \includegraphics[width=3.5in]{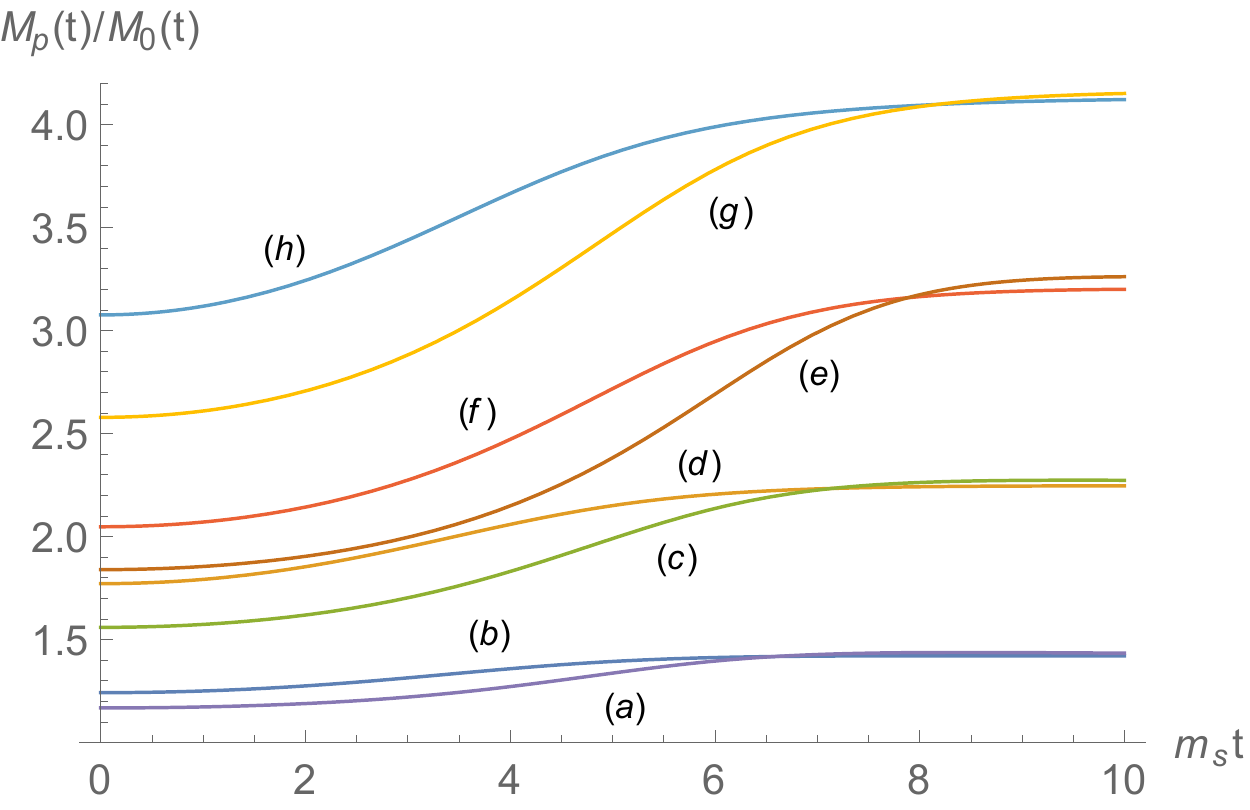}
  \end{center}
\caption{ \small (Color online) Ratio $M_p(t)/M_0(t)$ versus $\left(m_s t\right)$ for $0\leq m_s t \leq 10$, MDDs given by Eq. (\ref{Ofigs}), $\mu_0=m_s$, and different values of the parameter $\alpha$ and of the linear momentum $p$. Curve $(a)$ corresponds to $\alpha=1$ and $p= m_s$; $(b)$ corresponds to $\alpha=0$ and $p= m_s$; $(c)$ corresponds to $\alpha=1$ and $p=2 m_s$; $(d)$ corresponds to $\alpha=0$ and $p=2m_s$, $(e)$ corresponds to $\alpha=2$ and $p= 3 m_s$, $(f)$ corresponds to $\alpha=1$ and $p=3 m_s$; $(g)$ corresponds to $\alpha=1$ and $p=4 m_s$, $(h)$ corresponds to $\alpha=0$ and $p= 4 m_s$.}
\label{Fig13}
\end{figure}
\begin{figure}[!b]
  \begin{center}
    \includegraphics[width=3.5in]{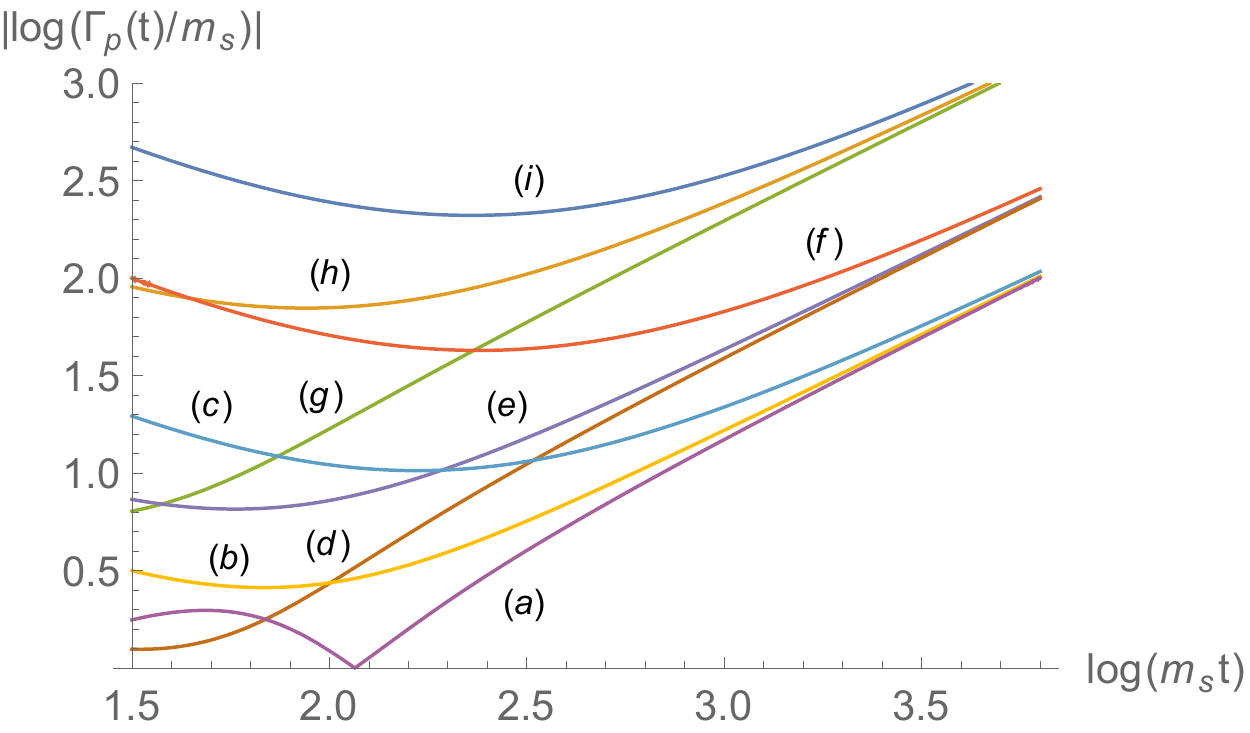}
  \end{center}
\caption{ \small (Color online) Quantity $\left|\log\left(\Gamma_p(t)/m_s\right)\right|$ versus $\log\left(m_s t\right)$ for $e^{1.5}\leq m_s t \leq e^{3.8}$, MDDs given by Eq. (\ref{Ofigs}), $\mu_0=m_s$, and different values of the parameter $\alpha$ and of the linear momentum $p$. Curve $(a)$ corresponds to $\alpha=2$ and $p=0 m_s$; $(b)$ corresponds to $\alpha=2$ and $p= 2 m_s$; $(c)$ corresponds to $\alpha=2$ and $p=4 m_s$; $(d)$ corresponds to $\alpha=1$ and $p=0m_s$; $(e)$ corresponds to $\alpha=1$ and $p=2 m_s$; $(f)$ corresponds to $\alpha=1$ and $p=5 m_s$; $(g)$ corresponds to $\alpha=0$ and $p=0 m_s$; $(h)$ corresponds to $\alpha=0$ and $p= 3 m_s$, $(i)$ corresponds to $\alpha=0$ and $p=5 m_s$.}
\label{Fig14}
\end{figure}
\begin{figure}[!b]
  \begin{center}
    \includegraphics[width=3.5in]{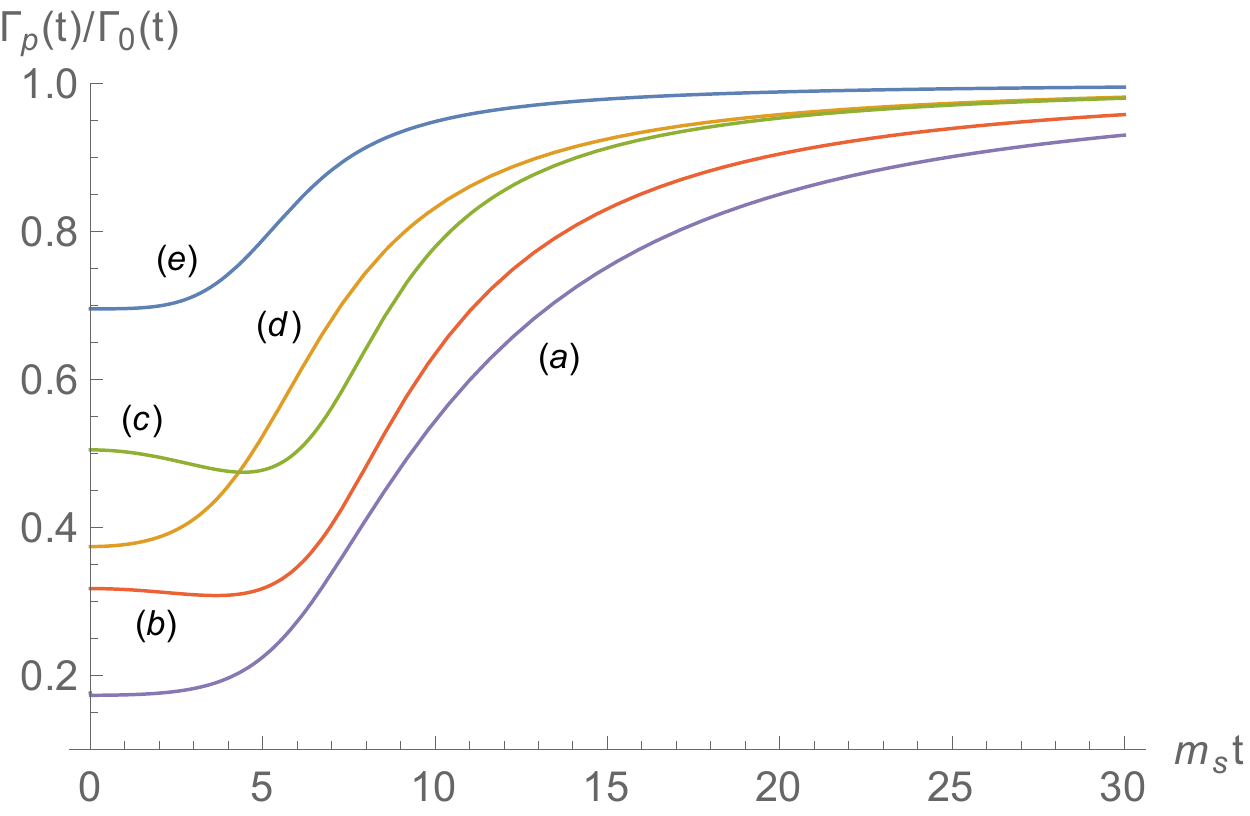}
  \end{center}
\caption{ \small (Color online) Ratio $\Gamma_p(t)/\Gamma_0(t)$ versus $\left(m_s t\right)$ for $0\leq m_s t \leq 30$, MDDs given by Eq. (\ref{Ofigs}), $\mu_0=m_s$, and different values of the parameter $\alpha$ and of the linear momentum $p$. Curve $(a)$ corresponds to $\alpha=1$ and $p=4 m_s$; $(b)$ corresponds to $\alpha=2$ and $p= 3 m_s$; $(c)$ corresponds to $\alpha=2$ and $p=2 m_s$; $(d)$ corresponds to $\alpha=0$ and $p=2m_s$, $(e)$ corresponds to $\alpha=0$ and $p=m_s$.}
\label{Fig15}
\end{figure}

The long-time behavior of the instantaneous mass and decay rate are studied in case the MDD fulfills the constraints which are reported in the second paragraph of Section 3. In addition, the functions $\xi \Omega_0\left(\xi\right)$ and $\xi \Omega\left(\xi\right)$ are required to obey the conditions which are requested in the same paragraph for the function $ \Omega_0\left(\xi\right)$ and the auxiliary function $ \Omega\left(\xi\right)$, respectively. The asymptotic analysis \cite{ErdelyiBook1956,WongBook1989} of the instantaneous mass or the instantaneous decay rate is obtained from 
the results of Section 3 and from Eq. (\ref{Epinst}) or (\ref{gEdef}), respectively.

The instantaneous mass tends over long times, $t \gg 1/m_s$, to the asymptotic value $M_p\left(\infty\right)$, given by 
\begin{eqnarray}
M_p\left(\infty\right)=\sqrt{\mu_0^2+p^2}, \label{Mpinfty}
\end{eqnarray}
according to the following dominant algebraic decay,
\begin{eqnarray}
M_p(t) \sim  M_p\left(\infty\right)\left(1+\zeta_p\left( m_s t \right)^{-2} \right). \label{MptLong}
\end{eqnarray}
The constant $\zeta_p$ is defined as
\begin{eqnarray}
\zeta_p= \left(1+\alpha\right)\frac{m_s}{\mu_0}\left(\left(1+\frac{\alpha}{2}\right)\frac{m_s}{\mu_0}\frac{p^2}{\mu_0^2+p^2}-\frac{\Omega^{\prime}_0\left(\mu_0/m_s\right)}{\Omega_0\left(\mu_0/m_s\right)}\right).\label{zetap}
\end{eqnarray}
For large values of the linear momentum, $p \gg\mu_0$, the instantaneous mass decays over long times, $t \gg 1/m_s$, as below,
\begin{eqnarray}
M_p(t) \sim p \left(1+\bar{\zeta}_{p}\left( m_s t \right)^{-2} \right), \label{MPtLong}
\end{eqnarray}
 where 
\begin{eqnarray}
\bar{\zeta}_p= \left(1+\alpha\right)\frac{m_s}{\mu_0}\left(\left(1+\frac{\alpha}{2}\right)\frac{m_s}{\mu_0}-\frac{\Omega^{\prime}_0\left(\mu_0/m_s\right)}{\Omega_0\left(\mu_0/m_s\right)}\right).\label{zetap}
\end{eqnarray}

If the linear momentum vanishes, $p=0$, the instantaneous mass (at rest) tends over long times, $t \gg 1/m_s$, to the minimum value of the mass spectrum,
\begin{eqnarray}
M_0\left(\infty\right)=\mu_0, \label{M0infty}
\end{eqnarray}
with the following dominant algebraic decay,
\begin{eqnarray}
M_0(t) \sim M_0\left(\infty\right) \left(1+\zeta_0\left( m_s t \right)^{-2} \right), \label{M0tLong}
\end{eqnarray}
 where 
\begin{eqnarray}
\zeta_0= -\left(1+\alpha\right)\frac{m_s}{\mu_0}\frac{\Omega^{\prime}_0\left(\mu_0/m_s\right)}{\Omega_0\left(\mu_0/m_s\right)}.\label{zeta0}
\end{eqnarray}

The instantaneous decay rate $\Gamma_p(t)$ vanishes over long times, $t \gg 1/m_s$, according to the following dominant algebraic decay,
\begin{eqnarray}
\Gamma_p(t) \sim \frac{ 2  \left( 1+ \alpha \right)}{t}.\label{GammaptLong}
\end{eqnarray}
Differently from the survival probability and from the instantaneous mass, the dominant asymptotic form of the instantaneous decay rate is independent of the linear momentum $p$. Consequently, the instantaneous decay rate at rest remains approximately unchanged over long times in the reference frame where the unstable particle moves with linear momentum $p$. Notice that the decay laws (\ref{M0tLong}) and (\ref{GammaptLong}) are in accordance with the ones which are obtained in \cite{GEPJD2016} for a wider class of MDDs.

Numerical analysis of the instantaneous mass or the instantaneous decay rate have been displayed in Figures \ref{Fig8}, \ref{Fig9}, \ref{Fig12}, \ref{Fig13}, or Figures \ref{Fig10}, \ref{Fig11}, \ref{Fig14}, \ref{Fig15}, respectively. The computed MDDs are given by the toy form (\ref{OmegaFig}) of the auxiliary function for different values of the non-negative power $\alpha$ and of the linear momentum $p$. 
The asymptotic lines of Figure \ref{Fig12} and the asymptotic horizontal lines of Figures \ref{Fig13} are in accordance with the long-time inverse-power-law decays of the instantaneous mass, given by Eq. (\ref{MptLong}) and Eq. (\ref{M0tLong}). The asymptotic horizontal lines of Figure \ref{Fig13} agree with Eq. (\ref{MpM0chip}) and with the expression (\ref{Chip}) of the scaling factor. The asymptotic lines appearing in Figure \ref{Fig14} and the asymptotic horizontal lines of Figures \ref{Fig15}, at ordinate $1$, agree with the long-time inverse-power-law behavior of the instantaneous decay rate, given by Eq. (\ref{GammaptLong}).

\section{Relativistic time dilation and survival probability}\label{6}

In Section 2.1 it is reported how the survival probability at rest, $\mathcal{P}_0\left(t\right)$, transforms, due to the relativistic time dilation, in the reference frame where the unstable particle moves with constant linear momentum $p$. The transformed survival probability, $\mathcal{P}_p\left(t\right)$, is related to the survival probability at rest, $\mathcal{P}_0\left(t\right)$, by the scaling law (\ref{ED}). The scaling factor consists in the corresponding relativistic Lorentz factor $\gamma_L$.

The analysis performed in Section 3 shows that, for the class of MDDs under study, the survival probability $\mathcal{P}_p\left(t\right)$ and the survival probability at rest $\mathcal{P}_0\left(t\right)$ are related by the scaling law (\ref{PpP0L}) over long times. The corresponding scaling factor $\chi_p$ is given by Eq. (\ref{Chip}). It is worth noticing that the scaling factor $\chi_p$ coincides with the ratio of the asymptotic form of the instantaneous mass $M_p(t)$ and the asymptotic expression of the instantaneous mass at rest $M_0(t)$ of the moving unstable particle,
\begin{eqnarray}
\chi_p=\frac{M_p\left(\infty\right)}{M_0\left(\infty\right)}.\label{MpM0chip}
\end{eqnarray}
At this stage, consider the reference frame $\mathfrak{S}$ where a mass at rest which is equal to the asymptotic value $M_0\left(\infty\right)$, becomes $M_p\left(\infty\right)$ due to the relativistic transformation of the mass. 
According to Eq. (\ref{MpM0chip}), in the reference frame $\mathfrak{S}$ the corresponding relativistic Lorentz factor coincides with the scaling factor $\chi_p$. 
We remind that, initially, the unstable quantum system is not in an eigenstate of the Hamiltonian. Consequently, in the present model the mass of the unstable particle is not defined. On the contrary, the instantaneous mass is properly defined in terms of the survival amplitude. See \cite{UrbAPB2017,UrbPLB2014,UrbanowskiAHEP2015} for details.

In light of the above observations, the long-time scaling law (\ref{PpP0L}) can be interpreted as an effect of the relativistic time dilation if the asymptotic value $M_p\left(\infty\right)$ of the instantaneous mass is considered to be the effective mass of the unstable particle over long times. In fact, in the reference frame $\mathfrak{S}$ the mass at rest $M_0\left(\infty\right)$, which is equal to the value $\mu_0$, moves with linear momentum $p$, or, equivalently, with constant velocity $1\Big/\sqrt{1+\mu_0^2/p^2}$, and becomes the relativistic mass $M_p\left(\infty\right)$, which is equal to the value $\sqrt{\mu_0^2+p^2}$. Concurrently, in the reference frame $\mathfrak{S}$ the survival probability at rest $\mathcal{P}_0\left(t\right)$ transforms, according to the relativistic time dilation, in the survival probability $\mathcal{P}_p\left(t\right)$ and obeys the scaling law (\ref{ED}), or, equivalently, (\ref{PpP0L}), over long times. In this context, the crucial condition of non-vanishing lower bound of the mass spectrum, $\mu_0/m_s>0$, suggests that the long-time relativistic dilation and the scaling law (\ref{ED}), or, equivalently, (\ref{PpP0L}), holds uniquely for an unstable moving particle with non-vanishing effective mass, $M_0\left(\infty\right)>0$.

\section{Summary and conclusions}\label{7}

The relativistic quantum decay laws of a moving unstable particle have been analyzed over short and long times for an arbitrary value $p$ of the (constant) linear momentum. The MDDs under study exhibit power-law behaviors near the (non-vanishing) lower bound $\mu_0$ of the mass spectrum. Due to the arbitrariness of the linear momentum, the ultrarelativistic and non-relativistic limits have been obtained as particular cases.

The survival probability, which is detected in the rest reference frame of the unstable particle, transforms in the reference frame where the unstable particle moves with linear momentum $p$, approximately according to a scaling law, over long times. The scaling factor is determined by the lower bound $\mu_0$ of the mass spectrum and by the linear momentum $p$ of the particle. The scaling law can be interpreted as the effect of the relativistic time dilation if the asymptotic form of the instantaneous mass $M_p(t)$ is accounted as the effective mass of the moving unstable particle over long times. In fact, consider the reference frame $\mathfrak{S}$ where a mass at rest of magnitude $\mu_0$ moves with velocity $1\Big/\sqrt{1-\mu_0^2\big/p^2}$, or, equivalently, with linear momentum $p$. The mass at rest $\mu_0$ coincides with the asymptotic value of the instantaneous mass at rest, $M_0\left(\infty\right)$, of the moving unstable particle. In the reference frame $\mathfrak{S}$ the transformed mass, which is equal to the value $\sqrt{\mu_0^2+p^2}$, coincides with the asymptotic value $M_p\left(\infty\right)$ of the instantaneous mass of the particle. Simultaneously, in the reference frame $\mathfrak{S}$ the dilation of times, which is suggested by the special relativity, transforms the survival probability at rest according to the mentioned scaling law. The above description indicates the value $1\Big/\sqrt{1-\mu_0^2\big/p^2}$ as the asymptotic velocity of the moving unstable particle.

We stress that the present interpretation is an attempt to ascribe the transformation laws of the long-time survival probability to the dilation of times which is provided by the theory of special relativity. However, a clear scaling transformation of the survival probability at rest holds, approximately over long times, if the decay is observed in the reference frame where the unstable particle moves with constant linear momentum. The scaling law can still be interpreted as the effect of a time dilation which appears by changing reference frame. The dilation is determined uniquely by the scaling factor which depends on the spectrum and on the dynamics of the unstable particle. The theoretical results are confirmed by the numerical analysis.

While the instantaneous mass transforms by changing reference frame, no transformation is found, approximately, for the instantaneous decay rate over long times. In fact, the instantaneous decay rate vanishes, over long times, approximately independently of the linear momentum of the moving particle. Consequently, the long-time instantaneous decay rate is approximately invariant by changing reference frame. In conclusion, the present analysis shows 
further ways to describe the long-time transformations of the decay laws of moving unstable particles in terms of model-independent properties of the mass spectrum. The role of the (non-vanishing) mass at rest in the relativistic transformation is assumed in the present description by the (non-vanishing) lower bound of the mass spectrum.



\begin{references}





\bibitem{Khalfin1997}
L.A. Khalfin, \emph{Quantum Theory of Unstable Particles and Relativity}, PDMI PREPRINT-6/1997, St. Petersburg Department of Stelkov Mathematical Institute, St. Ptersburg, Russia, 1997.





\bibitem{FondaGirardiRiminiRPP1978} L. Fonda, G.C. Ghirardi and A. Rimini, "`Decay theory of unstable quantum systems"', \emph{Reports on Progress in Physics}, vol. 41, no. 4, pp. 587--631, 1978.






\bibitem{BakamjianPR1961RQT}B. Bakamjian, "`Relativistic particle dynamics"', \emph{Physical Review}, vol. 121, pp. 1849--1851, 1961.


 \bibitem{CoePol1982RQT}F. Coester and W.N. Polyzou, "`Relativistic quantum mechanics of particles with direct interactions"', \emph{Physical Review D}, vol. 26, pp. 1348--1367, 1982.


\bibitem{ExnerPRD1983}P. Exner, 
"`Representations of the Poincar\'e group associated with unstable particles"', \emph{Physical Review D}, vol. 28, no. 10, pp. 2621--2627, 1983. 



\bibitem{StefBook2006}E.V. Stefanovich, \emph{Relativistic Quantum Theory of Particles}, vol. I and II, Lambert Academic, 2015.





\bibitem{HEP_Stef1996}E.V. Stefanovich, "`Quantum Effects in Relativistic Decays"', \emph{International Journal of Theoretical Physics}, vol. 35, no. 12, pp. 2539--2554, 1996.
\bibitem{HEP_Shir2004}M.I. Shirokov, "`Decay law of moving unstable particles"', \emph{International Journal of Theoretical Physics}, vol. 43, no. 6, pp. 1541--1552, 2004.



\bibitem{TD_FS1963} D.H. Frish and J.H. Smith, "`Measurement of the relativistic time dilation using $\mu$-mesons"', \emph{American Journal of Physics}, vol. 31, pp. 342--355, 1963.

\bibitem{TD_MuonsNat} J. Bailey, K. Borer, F. Combley, H. Drumm, F. Kreinen, F. Lange, E. Picasso, W. von Ruden, F.J.M. Farley, J.H. Field, W. Flegel, and P.M. Hattersley, "`Measurements of relativistic time dilatation for positive and negative muons in a circular orbit"', \emph{Nature}, vol. 268, pp. 301--305, 1977.

\bibitem{TD_Farley} F.J.M. Farley, "`The CERN (g-2) measurements."' \emph{Zeitschrift fur Physick C}, vol. 56, S88, Sect. 5, 1992.




\bibitem{HEP_Shir2006}M.I. Shirokov, "`Evolution in times of moving unstable systems"', \emph{Concepts of Physics}, vol. 3, no. 6, pp. 193--205, 2006.

\bibitem{HEP_Shir2009}M.I. Shirokov, "`Moving systems with speed-up evolution"', \emph{Physics of Particles and Nuclei Letters}, vol. 6, no. 1, pp. 14--17, 2009.


\bibitem{UrbPLB2014}
K. Urbanowski, "`Decay law of relativistic particles: Quantum theory meets special relativity"', \emph{Physics Letter B}, vol. 737, pp. 346--351, 2014.




\bibitem{TD_Giacosa2015} F. Giacosa, "`Decay law and time dilation"', \emph{Acta Physica Polonica B}, vol. 47, no. 9, pp. 2135--2150, 2016.


\bibitem{UrbanowskiAHEP2015}
K. Urbanowski, "`On the velocity of moving relativistic unstable quantum systems"', Advances in High Energy Physics. ID  461987, 6 pages, 2015.



\bibitem{UrbAPB2017}
K. Urbanowski, "`Non-classical behavior of moving relativistic unstable particles"', \emph{Acta Physica Polonica B}, vol. 48, no. 8, pp. 1411--1432, 2017.






\bibitem{GibPolBook}W.M. Gibson and B.R. Polard, \emph{Simmetry Principles in Elementary Particle Physics}, Cambridge, 1976.


\bibitem{BWMDD3} M. Goldberger and K. Watson, \emph{Collision Theory}, Wiley, New York, 1964.





   \bibitem{UrbanowskiEPJD2009}K. Urbanowski, "`General properties of the evolution of unstable states at long times"', \emph{European Physical Journal D}, vol. 54, no. 1, pp. 25--29, 2009.

\bibitem{UrbanowskiCEJP2009}
K. Urbanowski, "`Long time properties of the evolution of an unstable state"', \emph{Central European Journal of Physics}, vol. 7, pp. 696--703, 2009.

\bibitem{GEPJD2015} F. Giraldi, "`Logarithmic decays of unstable states"', \emph{European Physical Journal D}, vol. 69, article 5, 8 pages, 2015. 
\bibitem{GEPJD2016} F. Giraldi, "`Logarithmic decays of unstable states II"', \emph{European Physical Journal D}, vol. 70, article 229, 8 pages, 2016. 



\bibitem{UrbanowskiPRA1994}K. Urbanowski, "`Early-time properties of quantum evolution"', \emph{Physical Review A}, vol. 50, pp. 2847--2853, 1994.




\bibitem{Moller1972}C. M\o ller, The Theory of Relativity, Clarendon Press, Oxford, 1972





\bibitem{ErdelyiBook1956} A. Erd\'elyi, \emph{ Asymptotic expansions}, Dover, New York, 1956.
\bibitem{WongBook1989}R. Wong, \emph{ Asymptotic approximations of integrals}, Academic Press, Boston, 1989.


\end{references}
\end{document}